\documentclass{llncs}
\usepackage{makeidx}  
\usepackage{enumerate} 
\usepackage{graphicx} 
\usepackage{algorithm} 
\usepackage{algorithmic}
\usepackage{paralist}
\begin{document}

\newcommand{\bdaclass}{{\it bda-class}}
\newcommand{\bdaspace}{{\it bda-space}}
\newcommand{\bdaspaces}{{\it bda-spaces}}
\newcommand{\bdaheap}{{\it BDA-heap}}

\addtolength{\textheight}{3mm}

\title{Gang-GC: Locality-aware Parallel Data Placement Optimizations for Key-Value Storages}

\author{Duarte Patrício\inst{1,2} \and Jos\'e Sim\~ao\inst{1,3} \and Lu\'is Veiga\inst{1,2}}

\institute{
	INESC-ID Lisboa \and
	Instituto Superior T\'{e}cnico, ULisboa \and
	Instituto Superior de Engenharia de Lisboa (ISEL), IPL
}

\maketitle              

\begin{abstract}
	Many cloud applications rely on fast and non-relational storage to aid in the 
	processing of large amounts of data.
	Managed runtimes are now widely used to support the execution of several storage solutions of the NoSQL movement, particularly when dealing with big data key-value store-driven applications. The benefits of these runtimes can however be limited by modern parallel throughput-oriented GC algorithms,	where related objects have the potential to be dispersed in memory, either in the same or different generations. 
	In the long run this causes more page faults and degradation
	of locality on system-level memory caches.
	We propose, Gang-CG, an extension to modern heap layouts and to a parallel GC algorithm
	to promote locality between groups of related 
	objects.
	This is done without extensive profiling of the applications and in a way that
	is transparent to the programmer, without the need to use specialized data structures.
	The heap layout and algorithmic extensions were implemented over the Parallel Scavenge garbage collector
	of the HotSpot JVM\@. 
	Using microbenchmarks that capture the architecture of several key-value stores databases, we show negligible overhead in frequent operations such as the allocation of new objects and improvements to the access speed of data, supported by lower misses in system-level memory caches.
        Overall, we show a 6\% improvement in the average time of read
        and update operations and an average decrease of 12.4\% in
        page faults.
	
\end{abstract}

\section{Introduction}

The Java language is gaining space as the choice to
implement big-data processing and storage frameworks \cite{seven-databases,lakshman2010cassandra,Gidra0SSN15,Maas:2016}.
Java, and other bytecode-based languages, run on top of the Java Virtual Machine (JVM), and rely on it for just-in-time compilation and automatic memory management. This last sub-system is governed by the Garbage Collector (GC) which controls how object are allocated and when they are reclaimed from the heap. Although there are many tuning options for the JVM, and in particular regarding the GC, this sub-system can still be a cause of bottleneck for programs that rely on heavy usage of large memory spaces. This is mainly due to throughput oriented management mechanisms which can hinder co-locality of related objects and the way objects are represented and placed in memory \cite{Chen:2006:PPG:1133981.1134021,Bu:2013:BDB:2491894.2466485,Gidra:2013}. 




Space locality is known to have a relevant impact in performance \cite{Moon:1984:GCL:800055.802040,Wilson:1991:ELR:113446.113461,Gidra0SSN15}.
Wilson~\textit{et.\ al.\
}~\cite{Wilson:1991:ELR:113446.113461} presented reorganization
techniques to improve locality showing that there is improvement on
the GC's traversal algorithm when objects of a certain type are given
special treatment. 
Dynamic profiling was also studied so that information on frequency of access is gathered and used in the placement of those objects \cite{Chen:2006:PPG:1133981.1134021}.
And Ilham's work~\cite{ilham2011evaluation}
shows increased locality in system-level memory structures, such as
the L1D cache and the dTLB, when ordering schemes for children object
placement are accounted for, i.\ e.\@, Depth-First (DF), Breath-First
(BF) and Hot Depth-First (HDF). 
However, these works either apply a similar approach to all objects, which makes it difficult to tailor for storage-specify data-structures, or are hard to scale to very large heaps given the impact of per-object profiling in execution time. Furthermore they were not evaluated with modern parallel GC algorithms.

Our case study system, column and key-value databases, are primarily supported by associative structures. In the former case this is an internal aspect, while in the latter it is even expressed through public APIs. As the program progresses, insertion and removal of objects ends up placing apart what will be objects frequently accessed together. Objects are also scattered by the throughput-oriented nature of modern parallel garbage collections, which favour speed by copying bulks of memory regardless of the relation between them. 
This option causes lack of spacial locality and does not allow to take advantage of time locality, resulting in more page faults and misses on data and address translation caches. 
The typical organization of any object-oriented program, with a high level of delegation between data, gives a further contribution to the problem.

To reduce the impact of GC in the context of big data applications, others have made extensive modifications to the way certain objects are created and managed in special propose memory spaces, either requiring compiler and GC modifications or application-specific  data structures \cite{Bu:2013:BDB:2491894.2466485,NguyenWBFHX15}.

We take a different approach and make a reorganization to the way storage-relevant objects are promoted in the heap. We place together associative structures and their siblings, to increase space locality, but without disrupting other objects. This is done by modifications to the heap and a state-of-art GC algorithm, the Parallel Scavenge of the OpenJDK JVM, and without requiring the use of new data structures, making the solution easier to adopt in current and new systems.
The main contributions of this paper are:
\begin{enumerate}[i.]
	\item An extension for the Java heap to handle potential large
	objects of a certain type separately;
	\item A garbage collector extension which operate on such heap, divided
	into segments, promoting objects of the according types to these
	segments based on configurable decision criteria;
	\item Two approaches to save object's special status information, in
	order to reduce the amount of time it takes to read an object type
	and promote to the according space, i.\ e.\@, to minimize the
	collector's latency.
\end{enumerate}

The rest of the paper is organized as follows. Section \ref{sec:tech} presents the main building blocks of modern garbage collectors and what factors hinder locality in NoSQL big data storages. Section \ref{sec:bdaheap} presents modifications made to a modern parallel GC and heap organization to avoid the previous problems without modifications to the application. Section \ref{sec:eval} shows the small overheads of this solutions and the benefits at application and microbenchmark level. Section \ref{sec:concl} draws final conclusions.




\section{The case for locality improvements in NoSQL storage}
\label{sec:tech} 

Many of today's most used NoSQL databases are written in high-level languages \cite{seven-databases}, such as Java and C\#.
Doing so, developers rely on the services of managed runtimes, in particular the automatic memory management sub-system.
These runtimes use a shared address space, called \textit{heap}, which is
segmented during the initialization of the JVM according to the
garbage collector (GC) in use. 

Example of these storage systems include Cassandra~\cite{lakshman2010cassandra} and Oracle KVS \cite{oracle-kvs}. They are distributed,
column-oriented, ``NoSQL'' database, initially developed at
Facebook. The data model for Cassandra is a distributed key/value map
where the key is an identifier for the row and the value is a highly
structured object. 
These systems use large heaps to cache hot accessed data. It is however a challenge to efficiently manage such large allocation spaces.
In fact, when running the YCSB benchmark framework \cite{cooper2010benchmarking} with
five million entries in local mode, i.e., a load of 5 GBytes on
Cassandra, we noticed an excessive use of \textit{page swapping} and
garbage collection hanging the machine.
We address co-locality by investigating the promotion mechanism on the
garbage collector of manage runtimes.


\subsection{State-of-the-art in heap management}

Modern garbage collectors are generational, splitting the heap according to the age of the objects.
There can either be organized to run in concurrency with the threads belonging to the application, i.e. mutators, or stop the mutators and run a set of parallel threads to identify unreachable objects. The former are called \textit{concurrent} while the latter are named as \textit{parallel}.
We focus our attention on a widely used and state-of-the-art managed runtime, the OpenJDK Hotspot \cite{openjdk}. Here, the default GC for server-class machines is the Parallel Scavenge (PS) \cite{Gidra:2013}. Other options are the G1, a parallel algorithm, which uses a generational multi-segmented heap, and the CMS, a mostly concurrent mark and sweep algorithm. Studies show that, unless extensive tuning is made, the PS GC outperforms the other two options when using a large heap and taking into account different classes of big data applications \cite{DBLP:conf/bpoe/AwanBVA15,Gidra:2013}.
The PS GC divides the heap into two
generations, the \textit{young generation} and the \textit{old generation}. 
The young generation is further divided into three
areas, the \textit{eden}, and two \textit{survivor} spaces, \textit{to}
and \textit{from} spaces. A parallel copying collector operates on the
young generation and a two-phase mark-compact algorithm is used for
the old generation. 

In the young generation, PS uses a parallel copying algorithm in order to
promote surviving objects of the eden-space to the to-space or from
the from-space to the old generation, if the objects are past a specified
age threshold. 
A collection of this sort is also called a \textit{minor GC}.
If a young collection is not sufficient, then a \textit{full GC} takes
place. This operation is throughput-oriented because it does not move every
individual object but rather a batch of objects. 
When promoting, each GC thread moves live objects to thread-owned buffer called the {\it promotion-local allocation buffers} (PLAB).
Although it is a two-phase (mark and compact), in practice its work is divided in four
phases: 
\begin{inparaenum}
	\item the marking phase which marks all objects from the root-set;
	\item the summary phase which computes the destination address of each region;
	\item the compacting phase that drains and fills regions and the
	\item clean up phase to adjust certain pointers that cross region boundaries. 
\end{inparaenum}	
The marking phase and the compacting phase are executed in
parallel, whereas the summary phase and clean up phase are executed by
a single thread called the VM thread.

\subsection{Factors that hinder locality}

As datasets becomes larger, the multiple levels of object delegation results in objects that, although related, can be scattered across a large heap. 
Because PS is a throughput-oriented GC is does not improve co-locality
between dependent objects, such as those with a parent/child
relationship. This is so because the GC threads copy objects
independently as fast as possible, racing for the allocation of more PLABs. 
With this operation, and because there is usually several level of delegation to to reach data, the whole co-locality that could have been
present in the eden-space, i.e., at the object's creation and
initialization, is lost during promotion. This is especially
problematic for collections, such as arrays, lists and maps, when its
elements (the \textit{child}) are separated from the \textit{parent}
and from its \textit{siblings}.
It is our interest to keep certain collection elements concentrated on
a given location in the heap, so that the CPU can pull a
\textit{brotherhood} of elements into the same cache line and improve the use address translation caches. 

\section{Gang promotion in heap management}
\label{sec:bdaheap}

Dealing with the lack of locality between parent and children elements
requires modifications to the way objects are managed. To do so, we take into account the structure of a widely used GC algorithm, 
the Parallel Scavenge. The solution is however applicable to other generational GCs.
The new structure and algorithmic modifications are focused on the process of migration to the old generation, where several GB of data will be quickly placed.
We call it {\it gang promotion}\cite{gang-sched-2013} in relation to a technique called {\it gang scheduling}, a technique used by hypervisors to schedule related virtual CPUs and improve performance of dependent threads.

\begin{figure}[t]
	\centering
	\includegraphics[scale=0.6]{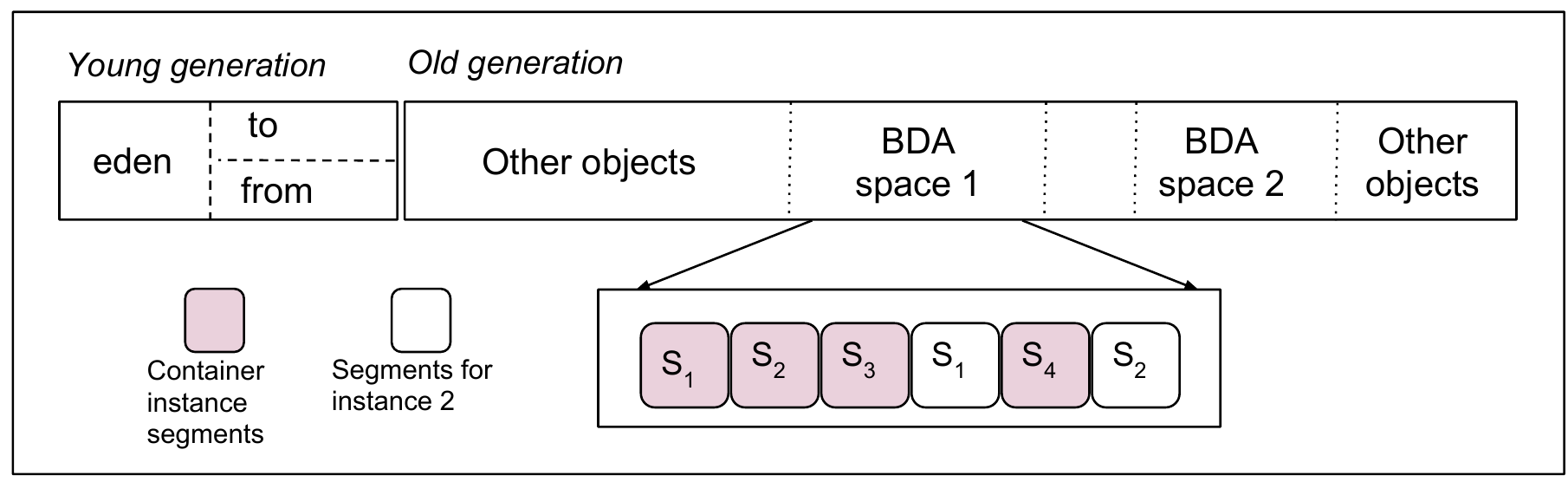}
	\caption{Partitioned old generation with two additional spaces}
	\label{fig:bdaheap}
\end{figure}

The new heap layout is termed \textit{BDA-heap}, short for ``Big Data
Aware Heap'', and is illustrated in Figure~\ref{fig:bdaheap}. 
The figure depicts the two generations and their internal logical divisions.
The \textit{BDA-heap} is organized in one or more {\it bda-spaces}. 
Each of these spaces stores instances of a given storage underlying type (e.g. \texttt{HashMap}, \texttt{SortedMap}) and their child-related objects, i.e.,  keys and data objects.
Inside each \textit{BDA-heap}, the objects belonging to a given map instance are organized into one or more memory {\it segments}, identified in Figure \ref{fig:bdaheap} as $S_x$. 
Therefore, affiliated segments, as presented with different shades in Figure \ref{fig:bdaheap}, correspond to a storage type instance and it is ensured that objects within the same segment are siblings. Segments for the same storage instance are grouped as a linked list. They can dynamically grow and are lazily allocated, allowing for the number of segments per instance to grow as more data is added. A group of related segments, not necessarily contiguous, is called a {\it container}.


\subsection{Identifying and managing space for storage instances}
\label{sec:identifyclass}
The relevant datatypes promoted and stored co-located with their child nodes, are hereafter termed as
\textit{bda-classes}. These classes need to be identified by the VM at runtime
during promotion so that the GC threads can redirect their instances and referenced objects to the correct {\it bda-space}. Given the high number of objects created in an
application it would be extremely inefficient to constantly scan the class of the object to check their relevance.

The {\it BDA-heap} uses a JVM-level reference queue which saves object
references and information regarding the target {\it bda-space}. An object is
enqueued when, during allocation, the VM checks that the type of the object being allocated corresponds to a class marked as relevant in the initial configuration parameters of the JVM. 
%

The {\it BDA-heap} layout is configured during launch time. Parameters include \bdaclass, the percentage of space used for \bdaspace~in regard to the whole old generation, and the number of bda-spaces to be managed. Storage objects refer to a group of objects which must be promoted in group. To guide this process a set of parameters are available, all optional except the \bdaclass:

\begin{itemize}
	\item \textbf{Classes} --- a list of comma-separated fully-qualified class names that the programmer wants the VM to handle as storage types.
	\item \textbf{BDA Ratio} --- the ratio of the old generation dedicated to the
	bda-spaces. This value is important since a small value can imply the
	allocation of segments in the non-bda-space;
	\item \textbf{Container Fraction} ($CF$) --- divide a container in smaller segments. This does not prevent the creation of more segments if need arises.
	\item \textbf{Delegation Level} ($DL$) --- the delegation level is an estimation of the level of delegation or indirection present in the objects referred by the storage type.
	\item \textbf{Default Number of Element fields} ($DNF$) --- data structures allocate, for each
	element, an object with one or more fields. This parameter sets the
	expected number of fields per each element;
      \item \textbf{Node fields} ($NF$) --- The parent data structure (e.g. a collection object) references its elements through an array or list of nodes, only then these nodes reference the elements. This parameter sets the number of fields in a node;
      \item \textbf{Container Size} ($CS$) --- This parameter sets the expected size for the container.
\end{itemize}

With the VM options presented, the algorithm can estimate how much a
collection of a certain type will occupy using Formula~\ref{eq:container_size}, where $h$ is the default header size for a given object
and $f$ the default field:

\begin{equation}
\label{eq:container_size}
(h + (NF) * f) * CS + DL * CS * (h + DNF * f)
\end{equation}


With Formula~\ref{eq:container_size}, the programmer can specify a
segment size, considering that he intends to fill the segment with
data, or else he will be fragmenting the heap. The programmer can use
the $CF$ option to divide into smaller segments decreasing
fragmentation, but it also increases GC times since there are more
segments to process.

\subsection{Gang GC phases}

This section discusses relevant changes made in the major phases of our target system, the Parallel Scavenge of OpenJDK.

\subsubsection{Young collection}
\label{sec:sub:ycoll}
When considering operations in the young generation, there are two main issues (1) the promotion of objects to the old
generation, and (2) the scan from the \textit{Old-To-Young} roots,
i.e., a part of the root-set consisting of objects in the old generation that reference young objects.

\paragraph{(1) Promoting to a bda-heap --}
Regular promotion of objects does not differentiate between root-references.
However, with a bda-heap, each GC thread grabs references
from the reference queue, presented in Section~\ref{sec:identifyclass},
allocates a new segment for each root (because objects in
the reference queue are container parents and not elements/children)
and pushes its fields, in depth-first, to a local stack where it will
later pop object references and promote them to the container of the parent.

\paragraph{(2) Scanning Old-To-Young --}
For the JVM to build the remember set, it instruments writes
performed by the mutator and dirties \textit{cards} in the
write-barrier card table, in a similar fashion as of H{\"o}lzle's method~\cite{holzle1993fast}.
When a range of dirty cards is found, it is adjusted to the start of the first object
and to the end of the last object in the range, pushing the object
fields into a queue for mark-through. 
But, these structures have no knowledge of boundaries in the middle of the
heap, because they know only the starting address of the 
space. Thus, in order to avoid scanning invalid addresses resultant of
unused space between each bda-space, our approach consists on scanning
each \bdaspace~independently, and, when possible, in parallel, 
preventing the traversal of its boundaries. 

\subsubsection{Full collection.}
\label{sec:sub:fcoll}
Because a full collection no longer promotes individual objects, but gangs of objects instead, we
detail how the 
existing algorithm was extended to cope with this,
concerning the four phases of the algorithm; (1) marking phase, (2)
summary phase, (3) compact phase and (4) cleanup phase. The order of
the paragraphs follows the same order that the collection steps take.

\paragraph{(1) Marking phase --}
In this phase, each parallel marking task marks the object-graph from the root-set, adds each
marked live object to the region that manages the
addresses they span and pushes their fields onto the marking stack to recursively handle all live objects.

\paragraph{(2) Summary phase --}
The summary phase, solely executed by the VM thread, computes the
destination for each region. It consists on iterating all regions that
contain live data and set the destination address of that region to the
last one set in the previous iteration, thus achieving compaction.
Since there is no way, performance-wise, to track which
objects in the young generation belong to one of the bda-spaces
containers, we do not drain young generation spaces to
bda-spaces. This allows us to keep the invariant that objects in a
container segment are related to others in that same
segment. However, since segments generate fragmentation, we devised a
summarize phase where we transfer segments of the same container to the bottom of the
corresponding \bdaspace, if there are available segments. Whenever we empty a segment, we return it to a pool and unlink
the previous owner, thus providing a fast means for another container to own it.

\paragraph{(3) Compact phase --}
Compaction is done in parallel such as marking. 
A region that contains work to do is one that can be
\textit{claimed} from the task queue.
The filling of a region consists in copying objects from the source
while they fit and, if the source was fully processed (it contains no
more words to copy) and there is still room for more, a new source is
fetched.
A new source region is
fetched by iterating through the adjacent source regions to find the
first that is not empty. Adjacent source regions can
target different container segments (no mixture of regions that span
different containers can happen), thus they cannot be fetched
unconditionally. This could result in the ``stealing'' of a region targeted
to another container, breaking our invariant. To avoid this, we have added a
simple condition to the algorithm that checks if the destination of a
new source region corresponds to the expected destination.

\paragraph{(4) Cleanup phase --}
The last phase for the Parallel Compact collector is to set the new
top pointers
delegating the call to the spaces themselves, clean the summary and
marking data, and clean or invalidate (set to dirty) the card
table barrier set depending whether the young generation is empty or
not, respectively. The \bdaheap~is transparent to the rest of the VM, 
therefore the top pointer of the old-space reflects the whole space occupied below. 
As a consequence, it is set to the top-pointer of the last bda-space. The dead space in
between bda-spaces is not problematic because whenever the VM tries to
access old-space, our virtual calls handle the access and direct the
VM to the correct segment. This is also the phase where we return
empty segments to the pool for later use by the corresponding container, reducing the process' working set.

\section{Evaluation}
\label{sec:eval}

Experimental runs were executed on a 4-core machine with 8
logical cores, three levels of cache with a 8MB L3 and 16GB of memory,
running a 64-bit Linux 4.4.0 kernel. 
Evaluation is split in two parts. The first one concerns the specific overhead of extra work during memory allocation. The second part shows how effective are the proposed modifications to the heap layout and gang promotion. This is done by analysing co-location of objects in memory and the speed of accesses to storage-related datatypes supported by higher ratios of cache hits.

\begin{figure}
	\centering
	\includegraphics[scale=0.5]{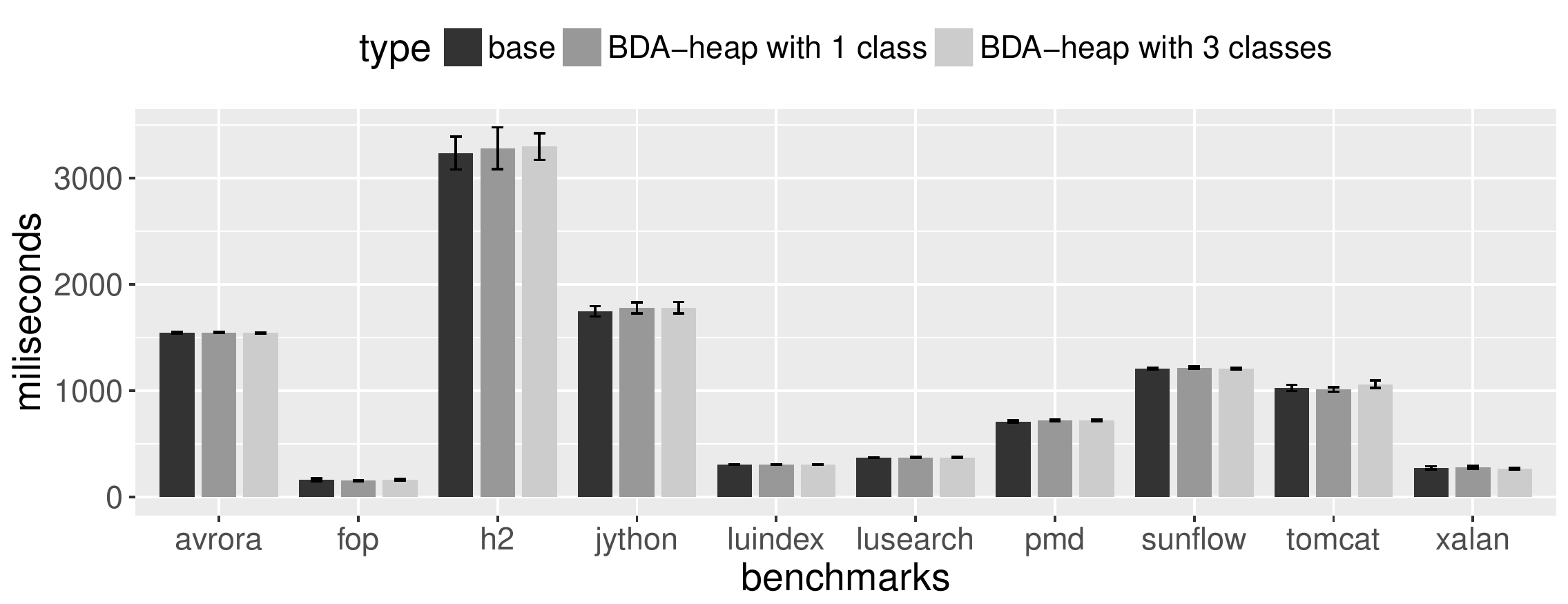}
	\caption{Overhead of additional verifications during object creation}
	\label{fig:callvm}
\end{figure}

To evaluate the impact of the reference queue, added to track new instances of bda-classes, we used the Dacapo benchmarks \cite{blackburn2006dacapo} which is heavy on allocations. Figure \ref{fig:callvm} shows the median execution time using the {\it base} OpenJDK distribution and our modified version while tracking 1 and 3 classes. Each benchmark was executed 10 times, with 10 warm-up iterations each. The results show a very small overhead in most benchmarks.

Key-value stores usually use multi-level map where, given a {\it table} name, a {\it row} name and a {\it field} name, a value can be inserted, read or updated \cite{seven-databases}:
\begin{equation}
\label{eq:maps}
\verb|map <table-name, map<row-key, sortedmap<field-key, data>>>|
\end{equation}
To evaluate the overall performance of the \bdaheap, we devised two micro-benchmarks that focus on this memory layout in different scenarios. 
The first one ({\it readonly}) creates several hash tables mapping a long value to a {\it string} with random values. Then, these maps were read (a \texttt{get(key)}) 200 million times per thread launched. The second one ({\it mapreduce}) is a map-reduce implementation using hash maps in the map and reduce phase, that first goes through a bootstrap phase where it generates several maps with random values in order to create enough pressure in memory. The map phase consists on grouping pairs of values with the same key and the reduce phase on grouping similar pair of values with the different keys across all maps.

The VM parameters used for these microbenchmarks were a mixed heap of 4 GBytes for the {\it readonly} and 6 GBytes for the {\it mapreduce}, and a $CS$ (see Section \ref{sec:identifyclass}) of 25k entries for both. 

\begin{figure}
	\centering
	\begin{minipage}{0.49\textwidth}
		\centering
		\includegraphics[width=1\linewidth]{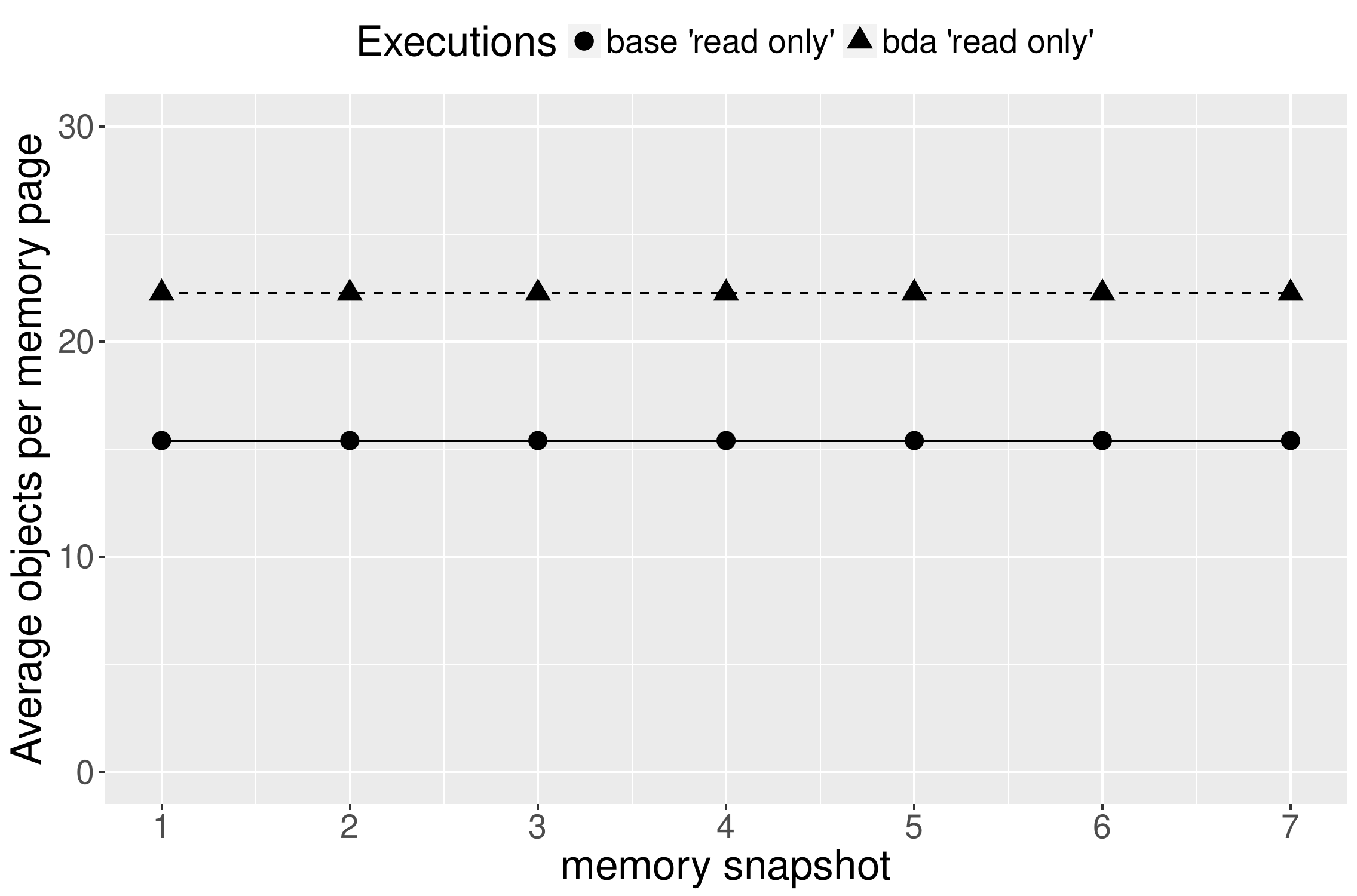}
		\caption{Average number of related objects per page in {\it read only} workload}
		\label{fig:objectspage-read}
	\end{minipage}
	\begin{minipage}{0.49\textwidth}
		\centering
		\includegraphics[width=1\linewidth]{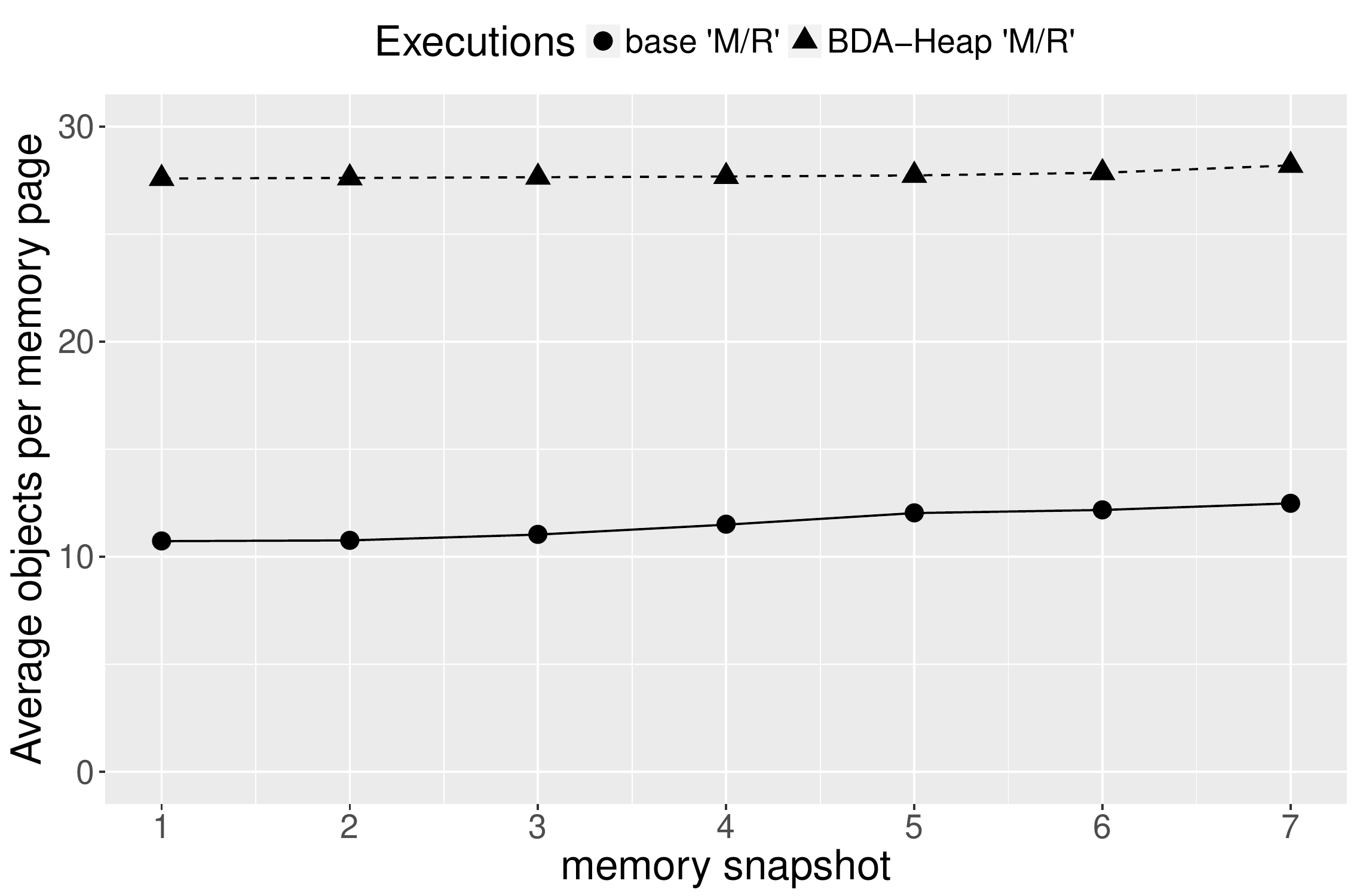}
		\caption{Average distance pointer between related object in {\it map and reduce} workload}
		\label{fig:objectspage-mr}
	\end{minipage}
\end{figure}

Figure \ref{fig:objectspage-read} and Figure \ref{fig:objectspage-mr} depict the average number of objects belonging to the same sub-graph per memory page in seven memory snapshots taken during execution. We resorted to a heap dump parser which, by previously coloring objects of the same subgraph, grouped these related objects and mapped each reference iteratively, taking into account address and page size/boundaries.

For the same workload, a higher number represents more objects per page, which means less pages for the same number of objects, contributing to less page faults and address translations. Figure \ref{fig:objectspage-read} shows an increase of 44\% in objects locality while in the case of map-reduce the number of objects per pages more than doubles when using a \bdaheap.

\begin{figure}
	\centering
	\begin{minipage}{0.49\textwidth}
		\centering
		\includegraphics[width=1\linewidth]{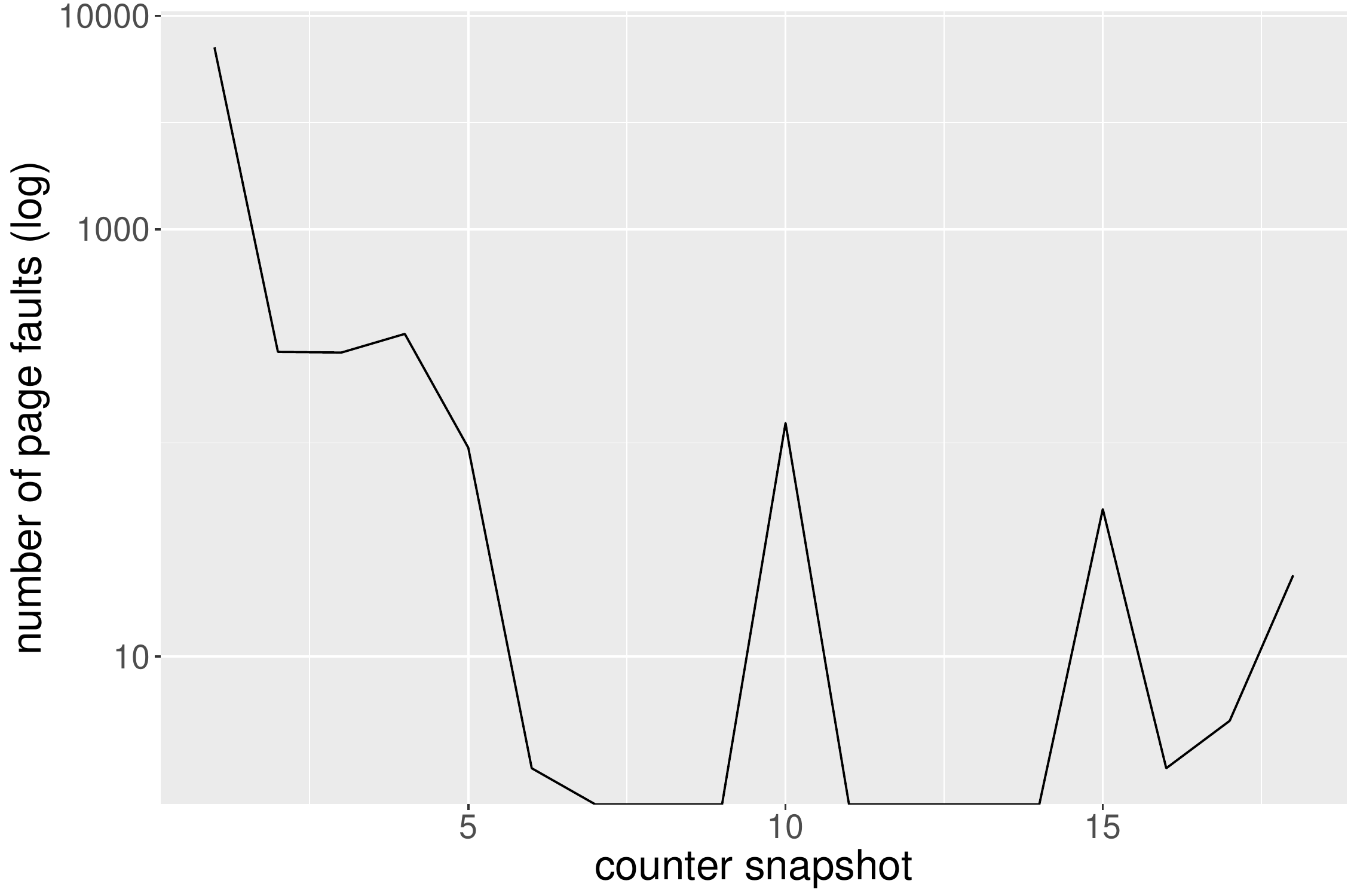}
		\\
		(a) base
	\end{minipage}
	\begin{minipage}{0.49\textwidth}
		\centering
		\includegraphics[width=1\linewidth]{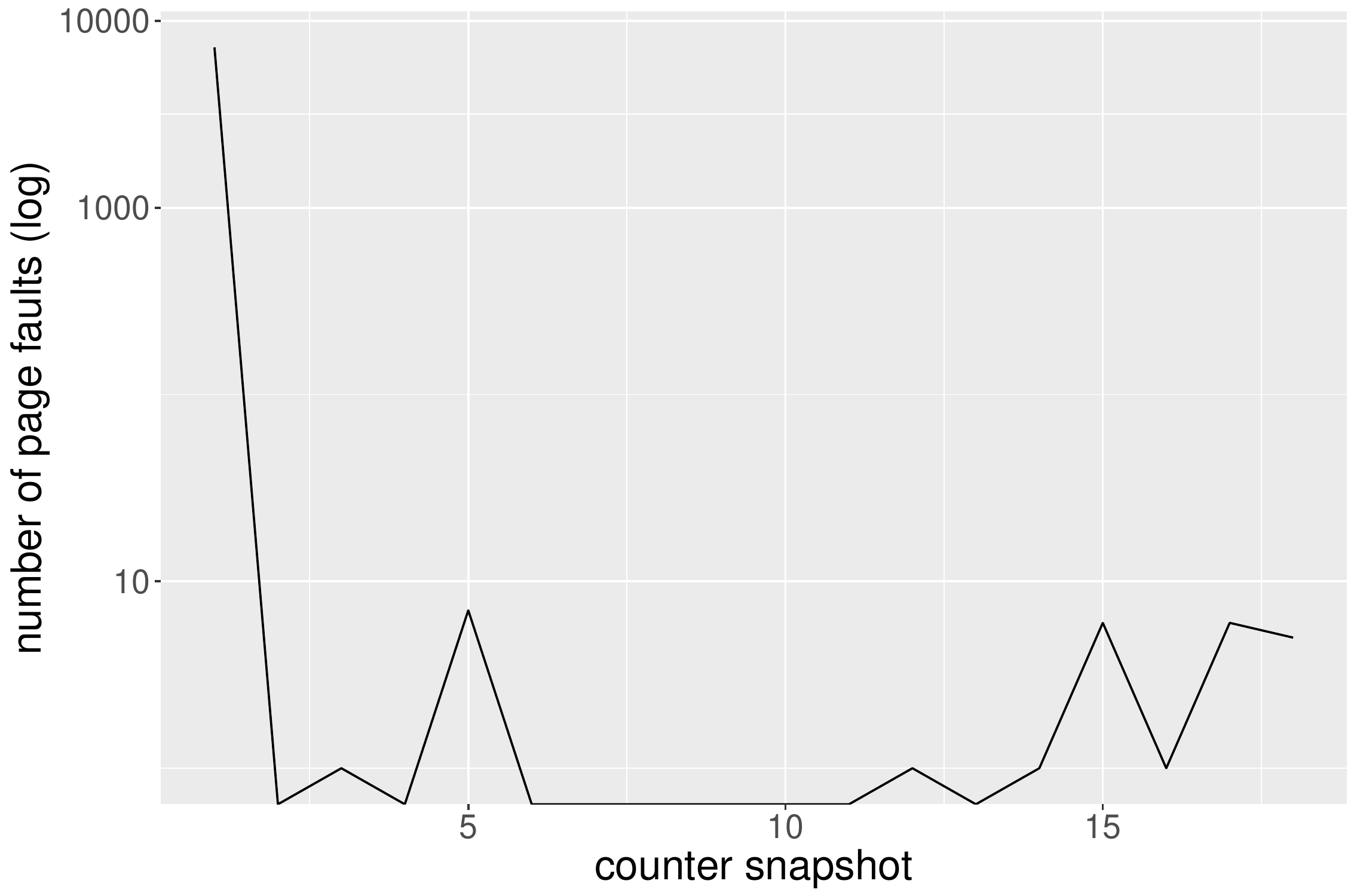}
		\\
		(b) BDA-heap
	\end{minipage}
	\caption{Page faults along the execution of map-reduce}
	\label{fig:page-faults}
\end{figure}

Figure \ref{fig:page-faults}.a) and Figure \ref{fig:page-faults}.b) show in the y-axis the number of page faults during the execution of the map-reduce micro-benchmark.
We can see improvements with the the BDA-heap organization, resulting in an average decrease of 12.4\% in page faults. Thus, just by employing a better suited GC algorithm (transparent to developers), we can improve in 6\% the average time of read, and also update, operations over the key-value data types used in this benchmark. This is possible because of the new organization of related objects and the gang promotion strategy.

\section{Related work}
\label{sec:relwork}

Research in automatic memory management has proven that there is no unique solution that fits all classes of applications. The best choice of GC is in many cases application and input-dependent \cite{Tay:2013}.
This has spanned a vast collection of algorithms, combining in many cases combinations of older ones, which can be stacked with application-specific profiles~\cite{singer2007intelligent}.

Parallel, stop-the-world algorithms have been making a successful entry in the field of big-data applications, since they
can efficiently collect a whole heap within shorter pauses and do not
require constant synchronization with the mutator, as it is the case with
concurrent collection \cite{Gidra:2013}. 
However, java-supported big data applications in general, and storage in particular, stresses the GC with lack of locality in large heaps and bloat of objects. This is mainly tackled using three kinds of approaches \cite{NguyenWBFHX15,Bu:2013:BDB:2491894.2466485,Maas:2016}: i) avoiding per-object headers and imposing new memory organizations at the framework-level, ii) speeding-up garbage collection by identifying objects that are created and destroyed together and, iii) coordinating the stop-the-world moment in inter-dependent JVM instances. Because most works focus on reducing overheads by dramatically changing the layout of objects and out-of-heap specially crafted structures, these solutions need changes both to the compiler and the GC system or rely on complex static analysis which is hard to prove correct and complete.

Facade ~\cite{NguyenWBFHX15} is a compiler and augmented runtime that reduces the number of objects in the heap by separating data (fields) from control (methods) and putting data in an off-heap structure without the need to maintain the bloat-causing header. Hyracks \cite{Bu:2013:BDB:2491894.2466485} is a graph processing framework that also uses a scheme where small objects are collapse into a special-propose data structures. Because this is done at the framework-level, and not the JVM-level, it is difficult to reuse the approach.
Overhead can also be caused by GC operations running uncoordinated inter-dependent JVM instances. When each of these instances need to collect unreachable objects, if it does so regardless of each other, this can cause pause significant pause times \cite{Maas:2016}.


On the other hand, previous work about object ordering schemes \cite{Moon:1984:GCL:800055.802040,Chen:2006:PPG:1133981.1134021} have shown that taking advantage of placement strategies, can increase locality in system-level memory structures and achieve better performance, especially when using guided techniques for optimal object placement. However, current approaches rely either on static analysis of fine-tunned dynamic profiling to avoid an excessive overhead. 


\section{Conclusions}
\label{sec:concl}

Several big data frameworks and storages are executed on a managed runtimes, 
taking advantage of parallel garbage collection and just-in-time compilation.
However, modern parallel memory management and throughput-oriented can hinder locality. Our approach was to promote objects' co-locality which minimizes the number of memory pages used, taking more advantage of system-level data and translation caches. This was done with an extension to the organization of a generation heap, called \bdaheap, and algorithmic modifications to a parallel GC, named as {\it Gang GC}.

The results provide positive conclusions about the inclusion of a
\bdaheap on a state-of-the-art JVM, the Hotspot, and the modifications
of the Parallel Scavenge to include the {\it Gang GC}
techniques. First, the DaCapo benchmarks showed that the extra work
required during memory allocation did not cause excessive
overhead. Second, the promotion efforts demonstrated a favourable
increase in page hits with micro-benchmarks of real world executions,
showing more object locality, less pages faults and,
consequently, less execution time per operation.
Future work include the evaluation with more specialized hardware, including NUMA
architectures, in order to stress our mechanisms further.



\bibliographystyle{splncs03}
\bibliography{references,references-suplements}

\end{document}